# On equations for neutrino propagation in matter

Paul M. Fishbane[*]
*Physics Dept. and Institute for Nuclear and Particle Physics,
Univ. of Virginia, Charlottesville, VA 22903*

Stephen G. Gasiorowicz[**]
*Physics Dept. and Institute of Theoretical Physics,
University of Minnesota. Minneapolis, MN 55455*

We study the dynamical equations for two-family neutrino oscillations in a medium of continuously-varying density. We can find explicit solutions to these equations in terms of series of nested integrals. These solutions can serve as a basis for numerical calculation of these processes or for further study of their analytical properties.

---

[*] email address pmf2r@virginia.edu
[**] email address gasior@umn.edu

## I. Introduction

The presence of neutrino oscillations [1-2] has renewed interest in the question of oscillations within matter [3-5]. The early work ("MSW") of Refs. 3 and 4 solved the problem of propagation within a medium of constant density, and it is possible to treat nonconstant density by numerical means. It is nevertheless always useful to think about an analytic approach [6] in order to develop insight and understanding. Since the case we study here is that of a two-channel problem, some aspects of the methods we describe are also applicable to spins in varying magnetic fields and other two-channel order-dependent problems.

For convenience we recall here the MSW results [3, 4], constant density We assume a two-channel approximation to neutrino mixing and give the amplitude $T(t)$ for a neutrino beam of energy $E$ passing through a medium of constant electron density $N_e$ given some initial neutrino flavor mixture $\psi(0)$, namely

$$\psi(t) = T(t)\psi(0). \tag{1.1}$$

(The time $t$ can be interchangably viewed as the thickness $x$ of the medium.) Here the two-vector weak state (i.e., flavor basis) is $\psi(t) = \begin{bmatrix} v_e \\ v_\mu \end{bmatrix}$. The transition amplitude $T$ is

$$T = \cos\phi' + i\sin\phi'\cos(2\theta')\sigma_z - i\sin\phi'\sin(2\theta')\sigma_x \tag{1.2}$$

where the primed variables contain the effect of the matter:

$$\delta m^2 \to \delta m'^2 = \delta m^2 \sqrt{\left(\cos 2\theta - \frac{2EA}{\delta m^2}\right)^2 + \sin^2 2\theta}$$

$$\varphi \to \varphi' = \frac{\delta m'^2}{4E} t \tag{1.3}$$

$$\theta \to \theta' \ni \sin^2 2\theta' = \frac{\sin^2 2\theta}{\left(\cos 2\theta - \frac{2EA}{\delta m^2}\right)^2 + \sin^2 2\theta}$$

with

$$A = \sqrt{2} G_F N_e. \tag{1.4}$$

The mass parameter $\delta m^2 = m_2^2 - m_1^2$, where $m_i^2$ is the $i^{th}$ mass eigenvalue, is positive. We recover the vacuum result, Cabibbo angle $\theta$, for $A = 0$.

This introduction sets some notation and recalls well-known results. Below we shall be concerned with variable density.

## II. Passage through a medium of variable density

The Schrödinger equation for a two-family weak state $\psi(t)$ propagating through a medium of electron density $N_e$ is

$$i\frac{d\psi}{dt} = (H_F + W)\psi, \tag{2.1}$$

where, using the approximation $m_i \ll E$ and after pieces in the hamiltonian proportional to the unit matrix are removed—they lead only to overall phases, as we describe further below—we have

$$H_F = \frac{\delta m^2}{4E} \begin{bmatrix} -\cos 2\theta & \sin 2\theta \\ \sin 2\theta & \cos 2\theta \end{bmatrix} \tag{2.2}$$

and

$$W = \begin{bmatrix} \sqrt{2} G_F N_e & 0 \\ 0 & 0 \end{bmatrix} \equiv \begin{bmatrix} A & 0 \\ 0 & 0 \end{bmatrix} \tag{2.3}$$

Here $H_F$ is related to the mass eigenstates by the Cabibbo matrix $V = \begin{bmatrix} \cos\theta & \sin\theta \\ -\sin\theta & \cos\theta \end{bmatrix}$,

$$H_F = V \left\{ \frac{1}{2E} \begin{bmatrix} m_1^2 & 0 \\ 0 & m_2^2 \end{bmatrix} \right\} V^\dagger \equiv V H_D V^\dagger. \tag{2.4}$$

A solution to Eq. (2.1) will give us the amplitude $T(t)$, through Eq. (1.1).

It is convenient to approach the dynamical equation (2.1) by starting in the mass basis, using the transformed states $\phi \equiv V^\dagger \psi$. We do so by multiplying Eq. (2.1) by $V^\dagger$, giving

$$i \frac{d\phi}{dt} = (H_D + V^\dagger W V) \phi. \tag{2.5}$$

Next we remove the factor $H_D$ by defining a new function $\xi$ by

$$\phi \equiv e^{-iH_D t} \xi.$$

Starting from Eq. (2.5) it is easy to see that this function obeys the equation

$$i \frac{d\xi}{dt} = e^{iH_D t} V^\dagger W V e^{-iH_D t} \xi. \tag{2.6}$$

In this expression, only the mass difference enters into the exponentials. To see this, define

$$\bar{E} \equiv E + \frac{m_1^2 + m_2^2}{4E} \quad \text{and} \quad \Delta \equiv -\frac{\delta m^2}{4E}. \tag{2.7}$$

In terms of these quantities, $H_D$ reads

$$H_D = \bar{E} + \sigma_3 \Delta$$

The portion of $H_D$ proportional to the unit matrix commutes through the quantities in Eq. (2.6) and cancels, leaving

$$i \frac{d\xi}{dt} = e^{i\Delta\sigma_3 t} V^\dagger W V e^{-i\Delta\sigma_3 t} \xi. \tag{2.8}$$

We next write out in $2 \times 2$ form the quantity multiplying $\xi$ on the right side of Eq. (2.8), using in particular the identity

$$e^{ia\sigma_3} = \cos a + i\sigma_3 \sin a = \begin{bmatrix} e^{ia} & 0 \\ 0 & e^{-ia} \end{bmatrix}. \tag{2.9}$$

We find

$$e^{i\Delta\sigma_3 t}V^\dagger W V e^{-i\Delta\sigma_3 t} = \frac{1}{2}A\begin{bmatrix} 1+\cos 2\theta & e^{2i\Delta t}\sin 2\theta \\ e^{-2i\Delta t}\sin 2\theta & 1-\cos 2\theta \end{bmatrix}$$
$$\to \frac{1}{2}A\begin{bmatrix} \cos 2\theta & e^{2i\Delta t}\sin 2\theta \\ e^{-2i\Delta t}\sin 2\theta & -\cos 2\theta \end{bmatrix} \quad (2.10)$$

In the last step we have taken out the piece proportional to the unit matrix; again, it contributes only an overall phase to $\xi$. (This can be seen in a variety of ways. For example, one can define a new function $\eta \equiv \exp\left(+\frac{i}{2}\int_0^t A(t')dt'\right)\xi$ and show that the equation for $\eta$ is identical to that for $\xi$ but with the last form in Eq. (2.10) on the right multiplying $\eta$.) We can finally transform away the term in Eq. (2.10) proportional to $\sigma_3$. We define the new function $\zeta$ by

$$\zeta \equiv \exp\left(+\frac{i}{2}\sigma_3 I(t)\cos 2\theta\right)\xi = \begin{bmatrix} \exp\left(\frac{i}{2}I(t)\cos 2\theta\right) & 0 \\ 0 & \exp\left(-\frac{i}{2}I(t)\cos 2\theta\right) \end{bmatrix}\xi, \quad (2.11)$$

where

$$I(t) \equiv \int_0^t A(t')dt'. \quad (2.12)$$

The function $\zeta$ obeys the simpler equation

$$i\frac{d\zeta}{dt} = e^{\frac{i}{2}\sigma_3 I\cos 2\theta}\begin{bmatrix} 0 & \frac{1}{2}Ae^{2i\Delta t}\sin 2\theta \\ \frac{1}{2}Ae^{-2i\Delta t}\sin 2\theta & 0 \end{bmatrix}e^{-\frac{i}{2}\sigma_3 I\cos 2\theta}\zeta$$

$$= \begin{bmatrix} 0 & \frac{1}{2}Ae^{i(2\Delta t+I\cos 2\theta)}\sin 2\theta \\ \frac{1}{2}Ae^{-i(2\Delta t+I\cos 2\theta)}\sin 2\theta & 0 \end{bmatrix}\zeta \quad (2.13)$$

$$\equiv P(t)\zeta.$$

The last line defines the matrix $P(t)$, in terms of which we can write our formal solution to this equation. We remark in particular, for later use, that the 21 element $P_{21}$ equals the complex conjugate of the 12 element $P_{12}$. We will below express the solution to Eq. (2.13) in terms of $P_{12}$ and $P_{12}^*$.

The form that $P(t)$ takes is easy to understand. The central matrix in the first line of Eq. (2.13) is a linear combination of $\sigma_1$ and $\sigma_2$, while the external factors $e^{\pm\frac{i}{2}\sigma_3 I\cos 2\theta}$ take the form of a rotation about the 3-axis. Their effect is then to rotate the combination of $\sigma_1$ and $\sigma_2$ to give a different combination that lies at a different angle—that takes the form of the original combination but with the phase shifted. This is indeed what happens, as the explicit calculations that give us $P(t)$ show.

**Solution of the equation for $\zeta$.** It is clear from the matrix structure of Eq. (2.13) that the commutator $[P(t), P(t')] \neq 0$, so that the solution to Eq. (2.13) involves an ordering. Formally, the solution is

$$\zeta(t) = \mathbf{T}\left[\exp\left(-i\int_0^t P(s)ds\right)\right]\zeta(0) \equiv M(t)\zeta(0), \tag{2.14a}$$

where the time ordering operator $\mathbf{T}$ ensures that the matrices $P$ in the expression are ordered so that $P$ with a later argument stands to the left of $P$ with an earlier argument. We may write

$$M(t) = \sum_{n=0}^{\infty} \frac{(-i)^n}{n!}\int_0^t ds_1 \int_0^t ds_2 \cdots \int_0^t ds_n \mathbf{T}\left[P(s_1)P(s_2)\cdots P(s_n)\right],$$

and it is a standard exercise to show that this is equivalent to

$$M(t) = \sum_{n=0}^{\infty} (-i)^n \int_0^t P(s_1)ds_1 \int_0^{s_1} P(s_2)ds_2 \cdots \int_0^{s_{n-1}} P(s_n)ds_n \tag{2.14b}$$

Finally we can separate out the explicit matrix elements of $M$ by using the matrix structure of $P$, namely $P(t) = P_x(t)\sigma_1 + P_y(t)\sigma_2$. This leads immediately to

$$M_{11} = 1 - \int_0^t P_{12}(t')\left(\int_0^{t'} P_{12}^*(t'')dt''\right)dt' + \ldots \tag{2.15a}$$

$$M_{12} = -i\int_0^t P_{12}(t')dt' + i\int_0^t P_{12}(t')\left(\int_0^{t'} P_{12}^*(t'')\left(\int_0^{t''} P_{12}(t''')dt'''\right)dt''\right)dt' - \ldots \tag{2.15b}$$

$$M_{21} = M_{12}\big|_{P_{12} \leftrightarrow P_{12}^*} \tag{2.15c}$$

$$M_{22} = M_{11}^* \tag{2.15d}$$

Alternatively, direct differentiation of the group of Eqs. (2.15) shows that it satisfies the necessary condition $\dfrac{dM}{dt} = -iPM$.

The solution given by Eqs. (2.15) is essentially a power series in $A$. The order dependence of the result is contained in the fact that the integrals in $M$ are nested. We can write the transition amplitude $T$ in terms of $M$ by undoing our series of transformations, leaving

$$T(t) = V\begin{bmatrix} e^{-i\left(\Delta t + \frac{1}{2}I(t)\cos 2\theta\right)} & 0 \\ 0 & e^{i\left(\Delta t + \frac{1}{2}I(t)\cos 2\theta\right)} \end{bmatrix} M(t)V^\dagger \tag{2.16}$$

Generally speaking the nested integrals in $M$ are complicated, even for the case of constant density. (Of course numerical integration is always possible.)

### III. Second order equation

Equation (2.13) represents two coupled first order differential equations. These equations can be rewritten as a single uncoupled second order equation, and the second order equation lends itself to a variety of treatments beyond the formal solution we have already expressed. In terms of the explicit components of $\zeta$, Eq. (2.13) reads

$$\begin{bmatrix} \dot{\zeta}_1 \\ \dot{\zeta}_2 \end{bmatrix} = \begin{bmatrix} 0 & \alpha \\ \beta & 0 \end{bmatrix} \begin{bmatrix} \zeta_1 \\ \zeta_2 \end{bmatrix} = \begin{bmatrix} \alpha \zeta_2 \\ \beta \zeta_1 \end{bmatrix}, \tag{3.1}$$

where

$$\alpha \equiv -iP_{12} \quad \text{and} \quad \beta \equiv -iP_{21} = -iP_{12}*. \tag{3.2}$$

One more derivative of, say, the upper component gives $\ddot{\zeta}_1 = \dot{\alpha}\zeta_2 + \alpha\dot{\zeta}_2 = \dfrac{\dot{\alpha}}{\alpha}\dot{\zeta}_1 + \alpha\beta\zeta_1$, or

$$\ddot{\zeta}_1 - \frac{\dot{\alpha}}{\alpha}\dot{\zeta}_1 - \alpha\beta\zeta_1 = 0 \tag{3.3}$$

Using the explicit expressions for $\alpha$ and $\beta$, we find that

$$\frac{\dot{\alpha}}{\alpha} = \frac{\dot{A}}{A} + i(2\Delta + A\cos 2\theta) \quad \text{and} \quad \alpha\beta = -\left(\frac{1}{2}A\sin 2\theta\right)^2.$$

Thus we have finally the second order equation

$$\ddot{\zeta}_1 - \frac{\dot{A}}{A}\dot{\zeta}_1 + i(2\Delta + A\cos 2\theta)\dot{\zeta}_1 + \left(\frac{1}{2}A\sin 2\theta\right)^2 \zeta_1 = 0 \tag{3.4}$$

It is worthwhile noting that the quantity $I$, Eq. (2.12), does not appear in this equation. We may take the required two boundary conditions to be $\zeta_1(0)$ and $\zeta_2(0) = \dot{\zeta}_1(0)/\alpha(0)$.

Once we find the solution for $\zeta_1(t)$, we have $\zeta_2(t) = \dfrac{1}{\alpha(t)}\dfrac{d\zeta_1(t)}{dt}$. We also remark here that we have verified that our formal solution to the equations for $\zeta$, Eq. (2.15), satisfies Eq. (3.4).

**Recovery of constant density case.** For the MSW case (constant $A \equiv A_0$), reviewed in Section I, Eq. (3.4) takes the form

$$\ddot{\zeta}_1 - ib_0\dot{\zeta}_1 + c_0^2\zeta_1 = 0 \tag{3.5}$$

where

$$b_o = 2\Delta + A_0\cos 2\theta \quad \text{and} \quad c_0 = \frac{1}{2}A_0\sin 2\theta. \tag{3.6}$$

If in addition we define

$$\omega \equiv \sqrt{b_0^2 + 4c_0^2}, \tag{3.7}$$

then the solution to this equation is

$$\zeta_1(t) = B_1\exp\left(\frac{it}{2}[\omega + b_0]\right) + C_1\exp\left(\frac{it}{2}[-\omega + b_0]\right) \tag{3.8a}$$

$$\zeta_2(t) = B_2\exp\left(\frac{it}{2}[\omega - b_0]\right) + C_2\exp\left(\frac{it}{2}[-\omega - b_0]\right), \tag{3.8b}$$

where

$$B_1 = \frac{1}{\omega}\left(\frac{1}{2}\zeta_1(0)[\omega - b_0] - c_0\zeta_2(0)\right)$$

$$C_1 = \frac{1}{\omega}\left(\frac{1}{2}\zeta_1(0)[\omega + b_0] + c_0\zeta_2(0)\right)$$

$$B_2 = \frac{1}{\omega}\left(\frac{1}{2}\zeta_2(0)[\omega + b_0] - c_0\zeta_1(0)\right)$$

$$C_2 = \frac{1}{\omega}\left(\frac{1}{2}\zeta_2(0)[\omega - b_0] + c_0\zeta_1(0)\right)$$

This solution allows us to identify $M(t)$ through Eq. (2.14) and hence the transition amplitude $T$ through Eq. (2.16) adapted to constant $A$, namely

$$T(t) = V\begin{bmatrix} e^{-\frac{it}{2}(2\Delta + A_0 \cos 2\theta)} & 0 \\ 0 & e^{\frac{it}{2}(2\Delta + A_0 \cos 2\theta)} \end{bmatrix} MV^\dagger = V\begin{bmatrix} e^{-\frac{it}{2}b_0} & 0 \\ 0 & e^{\frac{it}{2}b_0} \end{bmatrix} MV^\dagger \qquad (3.9)$$

The result of the exercise is

$$T = \frac{1}{2\omega}V\begin{bmatrix} (\omega - b_0)e^{i\omega t/2} + (\omega + b_0)e^{-i\omega t/2} & -2c_0\left(e^{i\omega t/2} - e^{-i\omega t/2}\right) \\ -2c_0\left(e^{i\omega t/2} - e^{-i\omega t/2}\right) & (\omega + b_0)e^{i\omega t/2} + (\omega - b_0)e^{-i\omega t/2} \end{bmatrix}V^\dagger \qquad (3.10)$$

Then specific calculation shows, for example,

$$T_{12}(=T_{e\mu}) = \frac{2i}{\omega}\Delta\sin 2\theta \sin(\omega t/2). \qquad (3.11)$$

This can be put into the canonical form of Ref. [4], $-i(\sin\varphi_m)(\sin 2\theta_m)$, where the subscript $m$ indicates the propagation is in material, and where $\varphi_m = \frac{\delta m^2_{eff}}{4E}t$, if we identify

$$\sin 2\theta_m = \frac{2|\Delta|\sin 2\theta}{\omega} = \frac{\sin 2\theta}{\sqrt{1 - \frac{4EA}{\delta m^2}\cos 2\theta + \left(\frac{2EA}{\delta m^2}\right)^2}} = \frac{\sin 2\theta}{\sqrt{\left(\cos 2\theta - \frac{2EA}{\delta m^2}\right)^2 + \sin^2 2\theta}} \qquad (3.12a)$$

and

$$\delta m^2_{eff} = 2E\omega = \delta m^2\sqrt{1 - \frac{4EA}{\delta m^2}\cos 2\theta + \left(\frac{2EA}{\delta m^2}\right)^2} = \delta m^2\sqrt{\left(\cos 2\theta - \frac{2EA}{\delta m^2}\right)^2 + \sin^2 2\theta}.\qquad(3.12b)$$

Indeed, in terms of these new variables the full transition amplitude is

$$T = \cos\varphi_m + (i\sin\varphi_m \cos 2\theta_m)\sigma_z - (i\sin\varphi_m \sin 2\theta_m)\sigma_x. \qquad (3.13)$$

This is the full canonical form described in Section I for propagation in a medium of constant density.

**Adiabatic Expansion.** If the factor in Eq. (3.4) that contains the derivative of $A$ is small compared to the other factors, one can make a systematic adiabatic expansion [6zz] in terms of it about the $0^{th}$ order (MSW) answer. To do so, it is useful to recast the solution technique somewhat. We shall first take the starting point of the neutrino beam, at $t = 0$, to specify the constant background level of the material density factor, i.e., $A(0) = A_0$. We leave the boundary conditions $\zeta_1(0)$ and $\zeta_2(0)$ unspecified for the moment but remark that using Eq. (2.13) the boundary condition for $\zeta_2(0)$ gives us alternatively a condition for the derivative of $\zeta_1$ at $t = 0$: $\dfrac{d\zeta_1}{dt}(0) = -iP_{12}(0)\zeta_2(0) = -\dfrac{i}{2}A_0(\sin 2\theta)\zeta_2(0)$.

We see from our earlier solution of the constant density case (Eq. 3.8) that the $t$-dependence is contained in a pair of phases. An alternative way to derive these phases in the constant density case is through a solution ansatz of the schematic form

$$\zeta_1 = R_0 \exp(iS_0(t)), \tag{3.14}$$

where $R_0$ is constant and where $S_0(t = 0) = 0$. The real and imaginary parts of Eq. (3.5) lead to the following equations for $S_0(t)$:

$$\left(\frac{dS_0}{dt}\right)^2 - b_0 \frac{dS_0}{dt} - c_0^2 = 0$$

$$\frac{d^2 S_0}{dt^2} = 0. \tag{3.15}$$

Together with the vanishing of $S_0$ at $t = 0$, these imply that

$$S_0(t) = \lambda t \tag{3.16}$$

where

$$\lambda^2 - b_0 \lambda - c_0^2 = 0 \tag{3.17}$$

The solution of the quadratic equation gives

$$\lambda_\pm = \frac{1}{2}(b_0 \pm \omega), \text{ where } \omega \equiv \sqrt{b_0^2 + 4c_0^2}. \tag{3.18}$$

Thus, as indeed Eq. (3.8) shows, a better ansatz for the solution is

$$\zeta_1(t) = R_+ e^{i\lambda_+ t} + R_- e^{i\lambda_- t} \tag{3.19}$$

The boundary conditions for $\zeta_1(t)$ give us immediately $R_+ + R_- = \zeta_1(0)$ and $R_+ \lambda_+ + R_- \lambda_- = -\dfrac{1}{2}A_0 \zeta_2(0)\sin 2\theta$; in turn these last two relations determine

$$R_\pm = \frac{\zeta_1(0)}{2} \mp \frac{b_0 \zeta_1(0) + A_0 \zeta_2(0)\sin 2\theta}{2\omega}. \tag{3.20}$$

Having reviewed the $0^{\text{th}}$ order (constant density) problem, we go on to include time (distance) dependence in the material density. We accordingly write the input density as

$$A(t) = A_0(1 + f_1(t)), \tag{3.21}$$

where $f_1(t) \ll 1$ for all $t$ in the problem and $f_1(t \leq 0) = 0$. We extend our ansatz for the solution to the form

$$\zeta_1(t) = R_+(t)\beta_+ e^{iS_+(t)} + R_-(t)\beta_- e^{iS_-(t)}, \tag{3.22a}$$

where
$$R_\pm(t) = R_\pm\left(1 + \rho_1^\pm(t)\right) \tag{3.22b}$$

$$S_\pm(t) = \lambda_\pm t + \sigma_1^\pm(t) \tag{3.22c}$$

The quantities with subscript "1" are all small; moreover, $\rho_1^\pm(0) = \sigma_1^\pm(0) = 0$. We also set
$$b(t) = b_0 + b_1(t), \text{ where } b_1(t) = A_0 f_1(t) \cos 2\theta \tag{3.23a}$$
$$c(t) = c_0 + c_1(t), \text{ where } c_1(t) = (A_0 f_1(t) \sin 2\theta)/2. \tag{3.23b}$$

We now insert our ansatz into Eq. (3.5). The spirit of the adiabatic expansion is to keep only first order terms in quantities with the subscript "1." In addition, we insist that the coefficients of $\exp(iS_\pm(t))$ vanish separately.

The real and imaginary parts of the coefficients of $\exp(iS_\pm)$ give respectively

$$\frac{d^2\rho_1^\pm}{dt^2} \mp \omega \frac{d\sigma_1^\pm}{dt} + K_\pm f_1 = 0 \tag{3.24a}$$

$$\frac{d^2\sigma_1^\pm}{dt^2} \pm \omega \frac{d\rho_1^\pm}{dt} - \lambda_\pm \frac{df_1}{dt} = 0, \tag{3.24b}$$

where

$$K_\pm \equiv A_0(\lambda_\pm \cos 2\theta + c_0 \sin 2\theta) = 2\lambda_\pm(-2\Delta \pm \omega)/2. \tag{3.25}$$

Equations (3.24) contain only derivatives of the functions we seek, so they are in fact two coupled first order equations for the functions

$$v_1^\pm \equiv \frac{d\sigma_1^\pm}{dt} \quad \text{and} \quad u_1^\pm \equiv \frac{d\rho_1^\pm}{dt}. \tag{3.26}$$

To the equations for $v_1^\pm$ and $u_1^\pm$ we add boundary conditions that follow from $d\zeta_1/dt = 0$, namely $v_1^\pm(0) = 0 = u_1^\pm(0)$. As we shall see, these boundary conditions guarantee that $\sigma_1$ and $\rho_1$ remain small (i.e., $O(f_1)$).

We decouple the two equations for $v_1^\pm$ and $u_1^\pm$ by taking one more derivative of, say, Eq. (3.24a), giving a single second order equation for $u_1^\pm$:

$$\frac{d^2 u_1^\pm}{dt^2} + \omega^2 u_1^\pm = 2\Delta\lambda_\pm \frac{df_1}{dt}, \tag{3.27}$$

where for the coefficient of $df_1/dt$ we have used $\pm\omega\lambda_\pm - K_\pm = 2\Delta$. This equation is solved by a standard Green's function $G(t - t')$ that satisfies

$$\frac{d^2 G(t-t')}{dt^2} + \omega^2 G(t-t') = \delta(t-t')$$

with, as causality suggests, $G(x) = 0$ for $x < 0$. The Green's function required is

$$G(x) = \frac{1}{\omega}\theta(t)\sin\omega t. \tag{3.28}$$

In terms of this function we have

$$u_1^\pm(t) = a_s^\pm \sin\omega t + a_c^\pm \cos\omega t + 2\Delta\lambda_\pm \int_{-\infty}^{\infty} dt' G(t-t')\frac{df_1(t')}{dt'}$$

$$= a_s^\pm \sin\omega t + a_c^\pm \cos\omega t + 2\Delta\lambda_\pm \frac{1}{\omega}\int_0^t dt' \frac{df_1(t')}{dt}\sin(\omega(t-t')) \tag{3.29}$$

(The lower limit reflects the fact that $f_1$ is identically zero for negative argument.) Once we have $u_1^\pm$ we can get $v_1^\pm$ from Eq. (3.24a):

$$v_1^\pm = \pm\frac{1}{\omega}\left(\frac{du_1^\pm}{dt} + K_\pm f_1\right)$$

$$= \pm\left(a_s^\pm \cos\omega t - a_c^\pm \sin\omega t + \frac{K_\pm}{\omega}f_1(t)\right) \tag{3.30}$$

At this point we can get $a_s^\pm$ and $a_c^\pm$ from the boundary conditions for $v_1^\pm(0)$ and $u_1^\pm(0)$, which determine $a_s^\pm = a_c^\pm = 0$. (It is worth noting that since these quantities are not proportional to $f_1$, the only way for them to be small is to be zero.) In turn, we have finally

$$\rho_1^\pm(t) = \frac{2\Delta\lambda_\pm}{\omega} \int_0^t \int_0^{t'} \sin(\omega(t'-t'')) \frac{df_1(t'')}{dt''} dt'' dt' \tag{3.31}$$

$$\sigma_1^\pm = \pm \frac{1}{\omega} K_\pm \int_0^t f_1(t') dt' \tag{3.32}$$

The phase function $S_\pm(t)$ takes on a suggestive form if we use the identity $K_\pm = \pm\omega\lambda_\pm - \Delta(b_0 \pm \omega)$, in which case we can write

$$S_\pm(t) = \lambda_\pm \int_0^t (1 + f_1(t')) dt' \mp \frac{2\Delta\lambda_\pm}{\omega} \int_0^t f_1(t') dt'. \tag{3.33}$$

The first term integrates the material density.

*An example*: We take a linear variation, $f_1 = qt$, together with the condition that the beam is pure $\nu_\mu$ at $t = 0$ (which translates into $\zeta_1(0) = -\sin\theta$ and $\zeta_2(0) = \cos\theta$). Then we have immediately

$$S_\pm(t) = \lambda_\pm \left(t + \frac{1}{2}qt^2\right) \mp \frac{2\Delta\lambda_\pm}{\omega} \frac{1}{2} qt^2 \tag{3.34a}$$

and

$$R_\pm(t) = R_\pm \left[1 + q\frac{2\Delta\lambda_\pm}{\omega^2}\left(t - \frac{1}{\omega}\sin\omega t\right)\right]. \tag{3.34b}$$

In Figs. 1 and 2 we plot some probabilities associated with the resulting amplitude for some representative values of the parameters.

## Acknowledgements

We would like to thank the Aspen Center for Physics, where much of this work was done. PMF would also like to thank Dominique Schiff and the members of the LPTHE at Université de Paris-Sud for their generous hospitality. This work is supported in part by the U.S. Department of Energy under grant number DE-FG02-97ER41027.

these citations is not, we believe, the one we have developed here. We shall discuss this matter elsewhere.

## Acknowledgements

We would like to thank the Aspen Center for Physics, where much of this work was done. PMF would also like to thank Dominique Schiff and the members of the LPTHE at Université de Paris-Sud for their hospitality. This work is supported in part by the U.S. Department of Energy under grant number DE-FG02-97ER41027.

## Figure Captions

Figure 1. Probability, as calculated in the adiabatic approximation described in the text, of $\nu_e$ as a function of time from production as a pure $\nu_\mu$ at $t = 0$, in a medium with density factor $A_0(1 + qt)$, where $A_0 = 6 \times 10^9$ cm$^{-1}$ = $10^{-13}$ eV corresponds to Earth-like density. We assume the primary mixing angle is $\theta = 0.7$ and that the difference of the square of the neutrino masses is $5 \times 10^{-6}$ eV$^2$. The factor $\Delta = -7.9 \times 10^{-14}$ eV, a value for which the neutrino energy lies around the MSW resonance value, an energy of roughly 20 MeV. The horizontal exis is in units of $10^{14}$ eV$^{-1}$; note that for $q = -2 \times 10^{-15}$ eV, $qt = -0.2$ at $t = 10^{14}$ eV$^{-1}$.

Figure 2. Same as Fig. 1, but with the factor $\Delta = -7.5 \times 10^{-15}$ eV, a value for which the neutrino energy lies roughly ten times higher than that corresponding to the MSW resonance value.

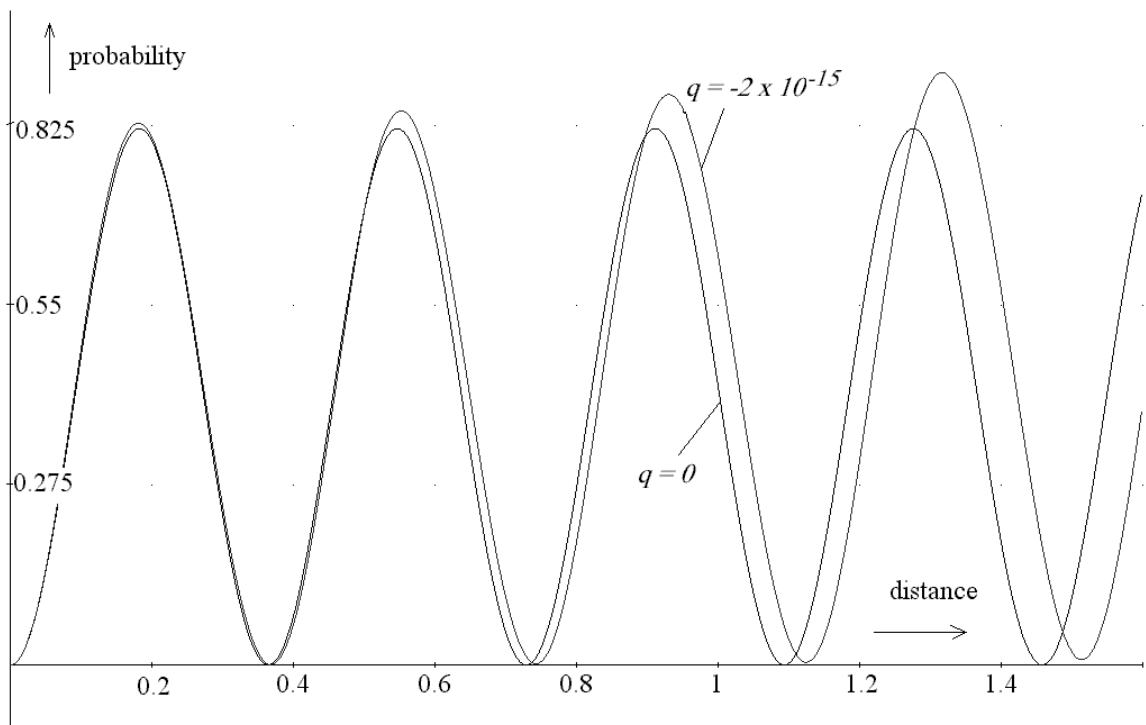

Figure 1

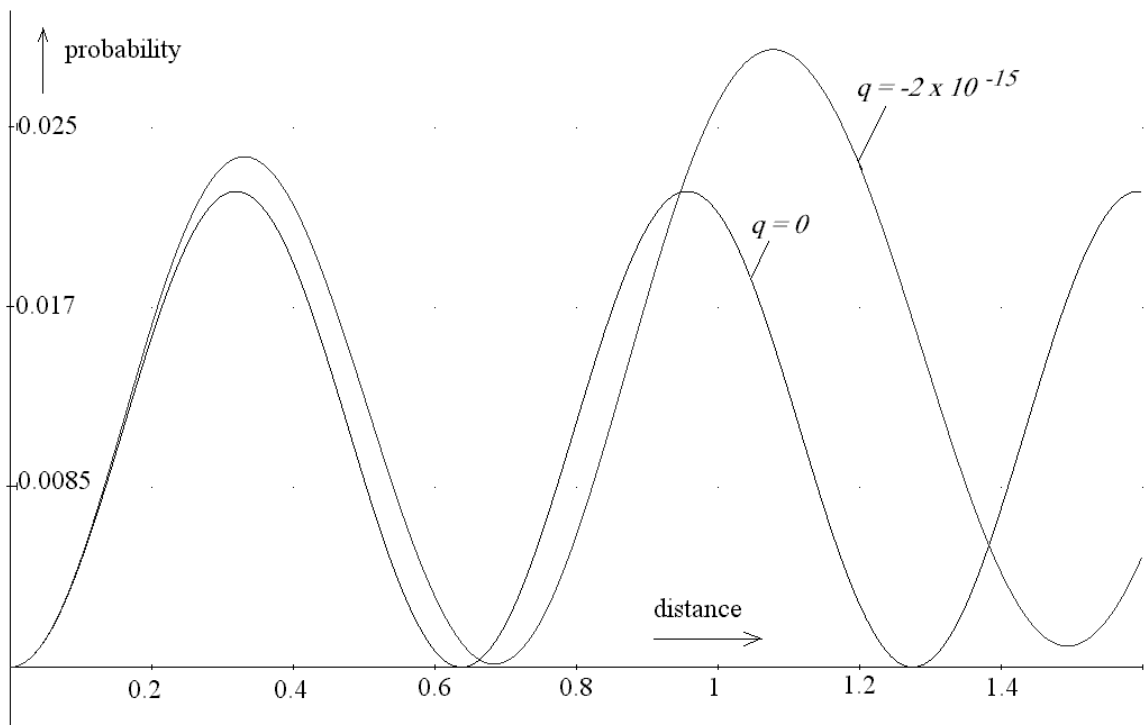

Figure 2